# Sensoring and Application of Multimodal Data for the Detection of Freezing of Gait in Parkinson's Disease

**Wei Zhang[1,3,†], Debin Huang[2,†], Hantao Li[2,†], Lipeng Wang[2], Yanzhao Wei[2], Kang Pan[2], Lin Ma[1], Huanhuan Feng[1], Jing Pan[1], Yuzhu Guo[2*]**

[1]Department of Neurology, Neurobiology and Geriatrics, Xuanwu Hospital of Capital Medical University, Beijing Institute of Geriatrics, Beijing 100053, China.
[2]Department of Automation Science and Electrical Engineering, Beihang University, Beijing 100191, China.
[3]Department of Neurology, The Affiliated Hospital of Xuzhou Medical University, Xuzhou, Jiangsu 221006, China.
†WZ, DH, and HL are equally contributed to this work.
*email addresses of corresponding authors: yuzhuguo@buaa.edu.cn

**Abstract:** The accurate and reliable detection or prediction of freezing of gaits (FOG) is important for fall prevention in Parkinson's Disease (PD) and studying the physiological transitions during the occurrence of FOG. Integrating both commercial and self-designed sensors, a protocal has been designed to acquire multimodal physical and physiological information during FOG, including gait acceleration (ACC), electroencephalogram (EEG), electromyogram (EMG), and skin conductance (SC). Two tasks were designed to trigger FOG, including gait initiation failure and FOG during walking. A total number of 12 PD patients completed the experiments and produced a total length of 3 hours and 42 minutes of valid data. The FOG episodes were labeled by two qualified physicians. Each unimodal data and combinations have been used to detect FOG. Results showed that multimodal data benefit the detection of FOG. Among unimodal data, EEG had better discriminative ability than ACC and EMG. However, the acqusition of EEG signals are more complicated. Multimodal motional and electrophysiological data can also be used to study the physiological transition process during the occurrence of FOG and provide personalised interventions.

## 1. Introduction

Parkinson's Disease (PD), with more than 10 million patients worldwide, is the second most prevalent brain malady after Alzheimer's disease [1]. Freezing of gaits (FOG), as one of the severest manifestations, grievously affects patients' life quality and may even menace the lives of aging patients. Therefore, accurate detection or prediction of FOG may significantly improve patients' life quality and promote personalized treatment of FOG.

FOG is an incapacitating issue, which refers to the interruption of the motion caused by the brain's incompetence to deal with concurrent cognitive and motor request signal input. A large quantity of research has revealed that FOG is accompanied by complicated variations of physiological signals. These signals are comprised of inertial acceleration data (ACC) of gaits, electroencephalogram (EEG), electromyogram (EMG), skin conductance (SC), and so on. A proper application of these data can make an accurate real-time detection or prediction of FOG. Data driven artificial intelligence methods to accurately detect or even predict FOG have attracted more and more attention. Handojoseno et al. studied the dynamic variation regularity of EEG signal during the occurrence of FOG and implemented detection and prediction of FOG using Bayesian neural networks (BNN) [2]. Mazilu et al. utilized SC to discriminate and anticipate FOG [3]. Cole et al. applied dynamic neural network (DNN) to ACC and EMG data to automatically detect FOG [4]. Deep learning (DL) methods proposed by Julià Camps et al. [5] provided a new prospective trend of FOG detection: either using traditional machine learning algorithms to extract features and then using DL algorithm to achieve accurate detection of FOG events or directly using end-to-end recurrent neural networks (RNN) to analyze time sequences and achieve FOG detection [6].

All these methods heavily depend on carefully designed experiments and reliable data. Specifically, although DL based methods provide a promising solution, they often need a big amount of data. The existing FOG dataset either with single modal physiological data or a limited number of participants due to the difficulties and complexities in the simultaneous acquisition of multimodal FOG data. This limits the detection accuracy of FOG and the generalization ability of the obtained detectors. A multimodal big data for FOG is desired for the detection or prediction of FOG. Acquisition of a sufficient amount of reliable data can be difficult because simultaneous high-precision acquisition of multiple signals requires complex experimental design and equipments. To the best of our knowledge, there is no public-available multimodal database that integrates ACC, EEG, EMG, and SC. In addition, such an intricate system that inevitably reduces the wearability of data acquisition system can significantly affect patients' actions and make the experiment deviate from the premise of studying FOG in patients' daily life settings. It is difficult to solve the dilemma of the stability of data and portability of sensors during walking tasks. Additionally, as FOG is sporadic, it is the experimental paradigm need to be framed to be able to engender FOGs. Therefore, it is crucial to build a software and hardware platform for multimodal data acquisition in Parkinson's patients with FOG.

Multimodal data provide comprehasive characterisation of the physiological process during the FOG and enable ones to reveal what information changes from non-FOG to FOG states, and vise versa. This can help us to figure out the physiological causality of FOG and study the inter-personal variabilities. In this study, multimodal data of 12 PD patients with FOG are acquired and analysed, including the expeimental design, sensoring system setup, data analysis, and detection of FOG based on multimodal data.

## 2. Materials and Methods

### *2.1 Participants*

In order to conduct the experiments safely and obtain valid data including sufficient FOG episodes, the participants were selected based on the following inclusion criteria:

1) Experiencing FOG during the off time;
2) Being able to walk independently during the off time;
3) No severe vision or hearing loss, dementia, or other neurological/orthopedic diseases.

### *2.2 Data Collection*

The multimodal sensoring platform acquires EEG, EMG, ACC, and SC. The locations of the sensors are shown in Fig 1.

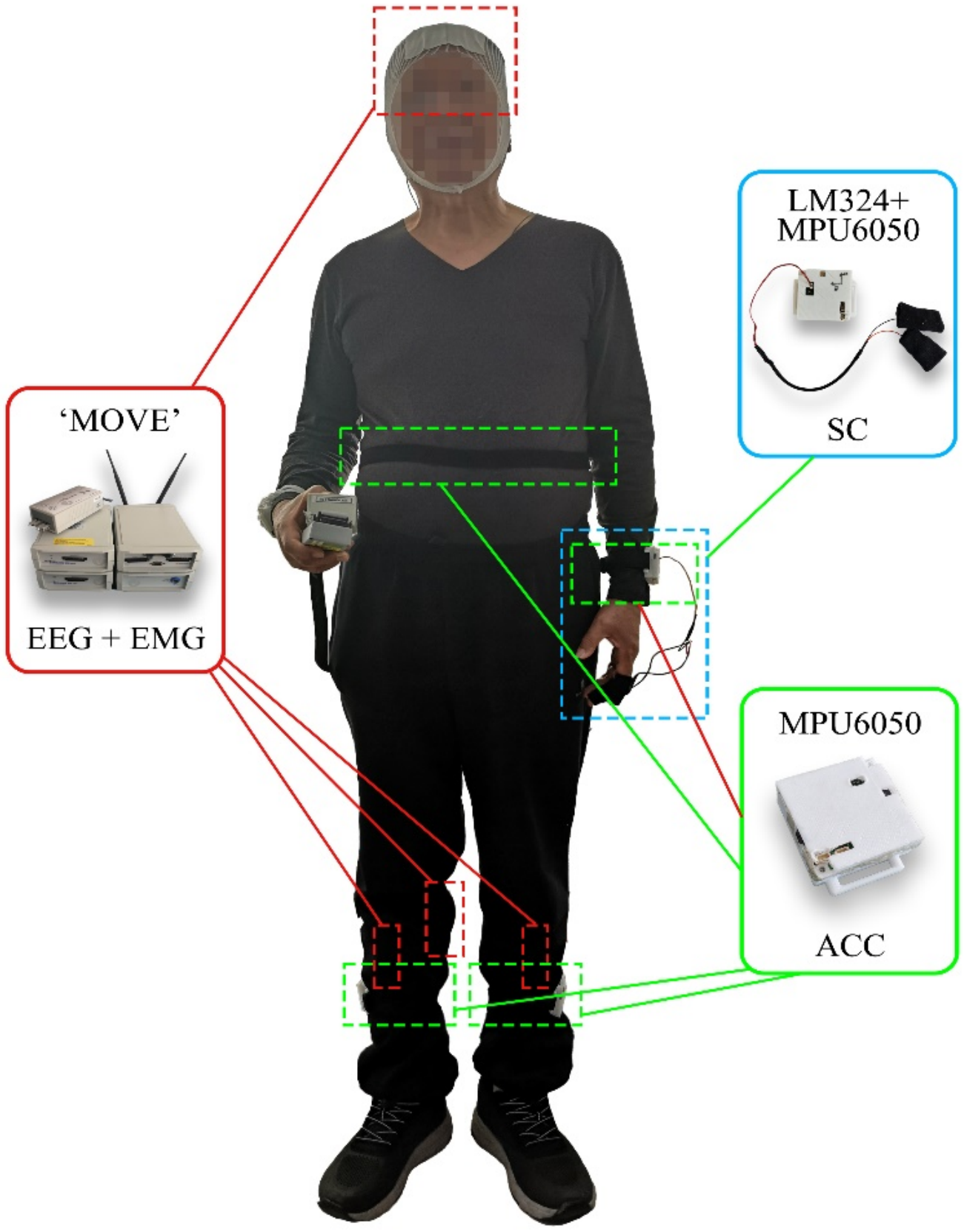

Fig. 1. The configuration of the FOG Multimodal data acquisition platform.

EEG and EMG were acquired using a 32-channel wireless MOVE system (BRAIN PRODUCTS, Germany). Among them, EEG signals at 28 channels (Fig 2) of the international 10-20 EEG electrode systems were recorded according to the results that FOG is related to the brain activities in the frontal, parietal, and occipital lobes [7]. The remaining 3 channels of the 'MOVE' system were used to collect EMG signals. The EMG signals were collected at the gastrocnemius muscle (GS) of the right leg and tibias anterior muscle (TA) of the left and right legs, respectively, according to the results in [8],which are shown in Fig 3.

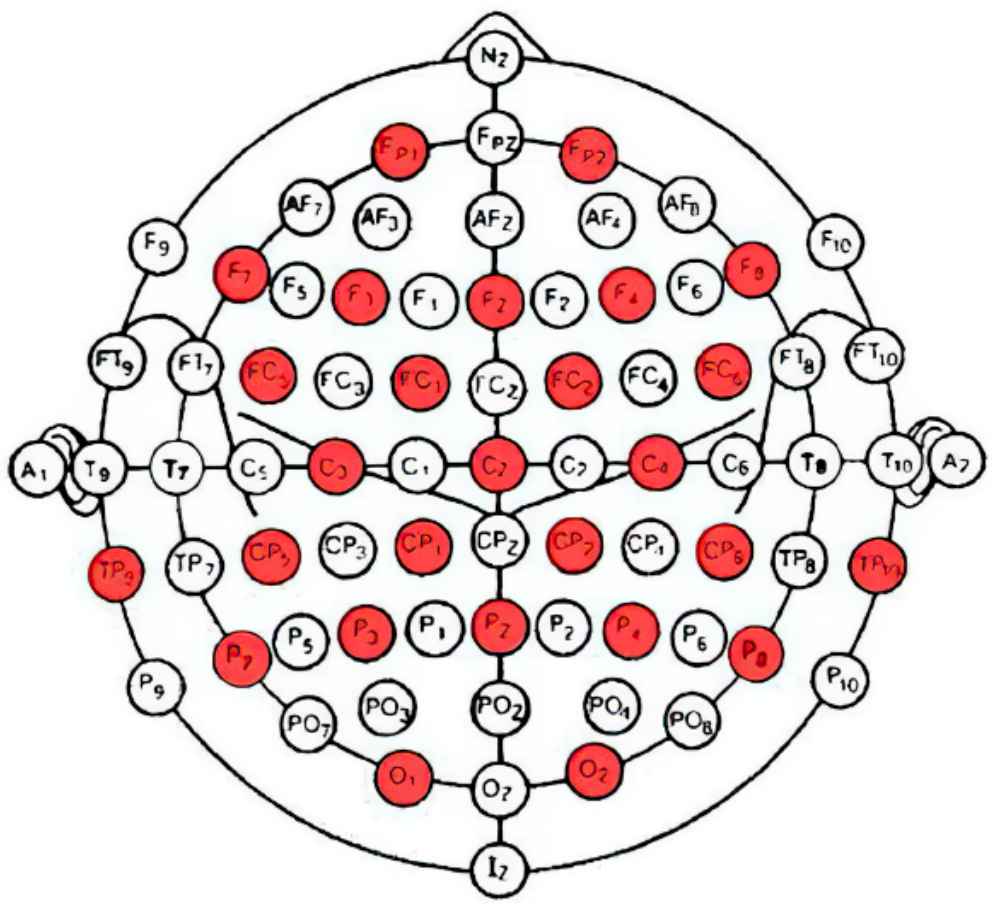

Fig. 2. The EEG channels recorded in the international 10-20 system.

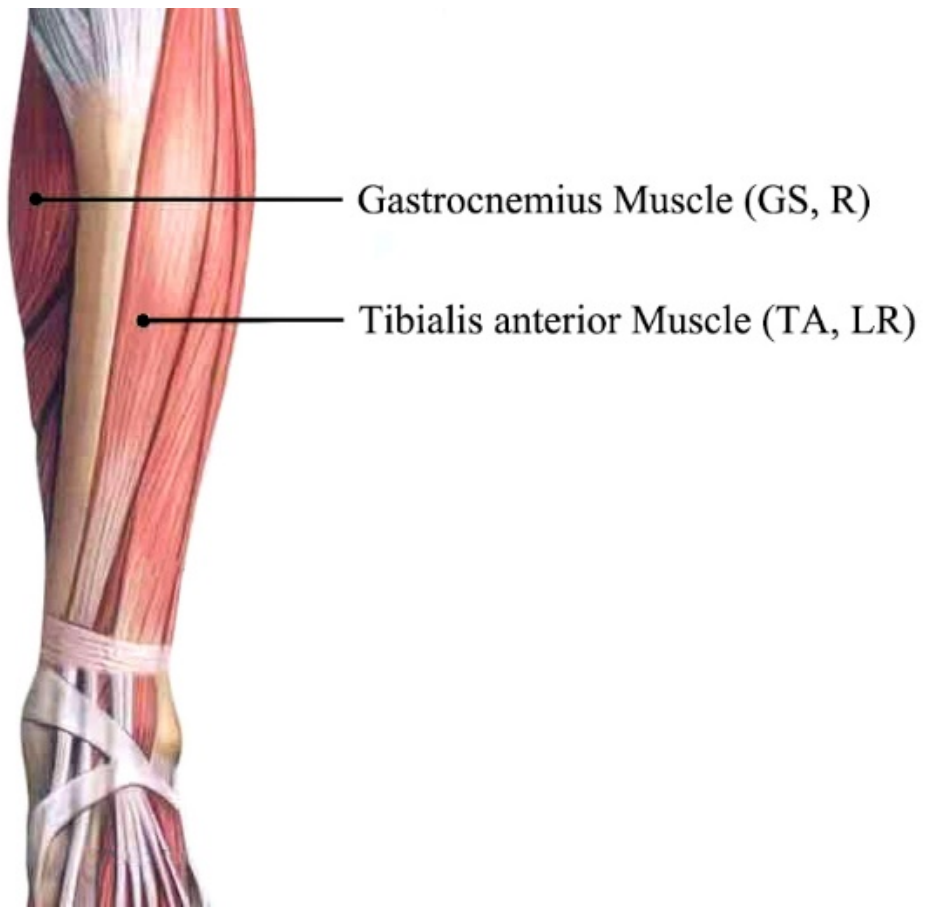

Fig. 3. Locations of the EMG sensor

ACC and SC were acquired using self-designed hardware subsystems based on TDK MPU6050 6-DoF accelerometer and gyro, with STMicroelectronics STM32 processor. Four accelerators were mounted at lateral tibia of the left and right legs, fifth lumbar spine (L5) of the waist and left arm, respectively. The SC acquisition was integrated in the ACC acquisition subsystem on the left arm. Both SC and ACCs were sampled at 500Hz and stored on a TF memory card. SC was recorded at the second belly of the left index finger and middle finger. The detailed information are sumarised in Table 1.

Table 1 Hardware configuration and location of the sensoring system

| Sensing Type | System | Sensor Quantity | Sensor Location |
|---|---|---|---|
| *28D*-EEG | *'The wireless MOVE'* | 28 | FP1, FP2, F3, F4, C3, C4, P3, P4, O1, O2, F7, F8, P7, P8, Fz, Cz, Pz, FC1, FC2, CP1, CP2, FC5, FC6, CP5, CP6, TP9, TP10, IO |
| *3D*-EMG | | 3 | Gastrocnemius muscle of right leg; Tibialis anterior muscle of left and right legs |
| *3D*-accelerometer | MPU6050 | 4 | Lateral tibia of left and right legs; Fifth lumbar spine; Wrist |
| *3D*-Gyro | | 4 | |
| *1D*-SC | LM324 | 2 | The second belly of the index finger and middle finger of the left hand |

Some patients only wear two inertial sensors (mounted on left tibia and left wrist, respectively).
TP9, TP10 (signal of the mastoid process of the temporary bone) were used as reference in data preprocessing.
IO (electrooculogram) was given in the dataset without preprocessing.

*2.3 Protocol*

The experiments were conducted in Beijing Xuanwu Hospital, China. Ethical approval (No. 2019-014) was obtained from Ethics Committee of Xuanwu Hospital, Capital Medical University, Beijing, China, and the research was conducted according to the declaration of Helsinki. Written informed consents were obtained from all participants. Data were collected in the off-medication state of patients.

FOG can be affected by environments and patient's emotional states and often happen in living circumstance. Tasks which may trigger FOG have been well reported in the literature, including walking through narrow spaces, approaching obstacles, turning and so on [9]. Based on this knowledge, the experimental proceduer was designed to include two types of tasks to trigger FOG episodes.

The procedures of the exepirment is given as follows:

1) Participants read and sign the informed consent;

2) Participants are asked to take a physical examination and fill in medical history form and Unified PD Rating Scale (UPDRS) questionnaires to confirm participants meet the inclusion criteria. Participants do not take medicine within 2 hours before collecting data to ensure that they are in off time;

3) Participants wear the multimodal sensing equipment with the help of professional technicians. The EEG cap, EMG electrodes, ACC, and SC sensors are mounted at the specified locations as shown in Fig 1 - 3 and Table 1;

4) Complete tasks 1-4 which are defined as follows according to the experimental paradigm. Video is recorded during the whole experiment for labelling FOG and non-FOG intervals;

5) Check the saved data at the end of the experiment to confirm no faults in the process. Otherwise, discard the data which do not meet the requirements and re-do the fault task after a 2-minute rest.

Each participant conducted four walking tasks, which were defined as:

1) Task1: The walking task was conducted in a setting as shown in Fig.4. Participants started from a sitting state. When a participant is ready, they rise from a chair at Point A and marches to the junction B between the room and a narrow corridor. Turn right and walk into the corridor. Bypass the obstacle 1 (can be a chair or a square region on the floor) by turning their bodies. Continue going straight along the narrow corridor until Point C. Make a U-turn at the end of the corridor, and go along an opposite direction. Bypass the three obstacles 1,2 and 3 by turning their body. When they reach the left end of the corridor, Point D, make another U-turn, bypassing obstacles 3 and 2, and reach the door of the room. Enter the room, and walk back to the chair, and sit down.

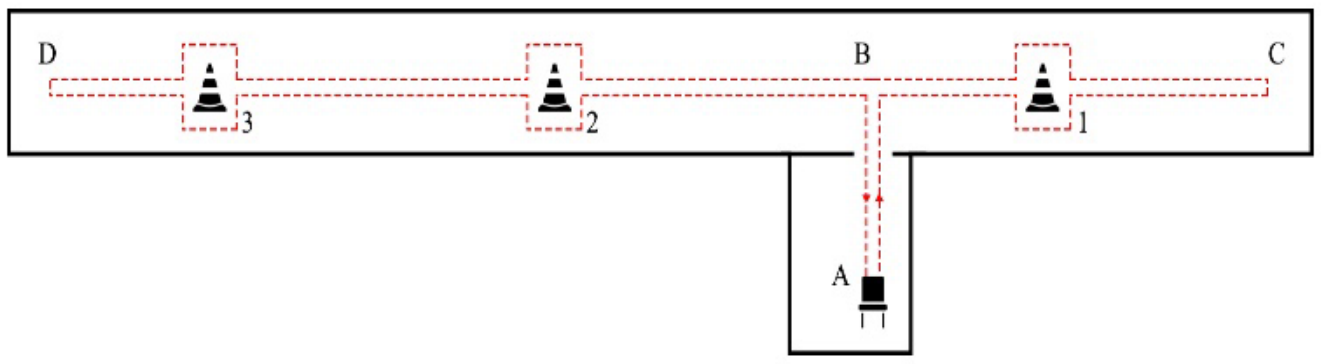

Fig. 4. Experimental settings of task 1 and 2.

2) Task 2: Repeat Task 1 one more time.

3) Task 3: This task was conducted in a setting as shown in Fig.5, where a square was drawn on the ground for patients to make a turn in a limited space. When the patient is ready, stand up from the chair at the end A in the room and march to the pre-pasted square mark at the end B in the room. The participant makes a U-turn in the narrow square region, and then walk straight back to the chair and sit down.

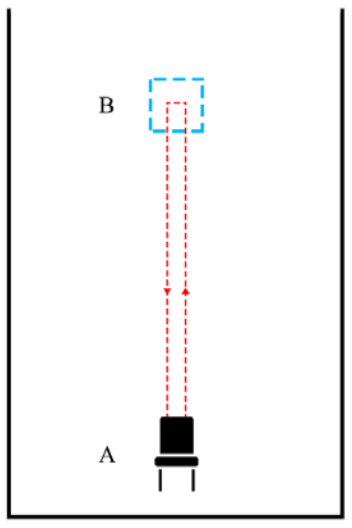

Fig. 5. Experimental settings of task 3 and 4.

4) Task 4: Repeat Task 3. Then end the experiment.

*2.4 Data Pre-processing*

Since the multimodal sensors recorded data separately time alignment of the multimodal data is essential for the consecutive multimodal data analysis and their application in FOG detection. The multimodal data were aligned based on time stamps generated by individual sensing subsystems. Firstly, data of different sampling frequencies will be resampled to a unified sampling rate of 500Hz. That is, the EEG and EMG which were sampled at 1000Hz were down-sampled to 500Hz that was the sampling frequcy of ACC and SC subsystems. A cubic interpolation method were used to calibrate all the singles to the time stamps of ACC subsystems. The rules of data alignment were given in Appendix A.

Two qualified physicians from the Department of Neurology, Beijing Xuanwu Hospital labeled the time instances when a FOG started and ended in the video, respectively. The labels were assigned to each data points in the aligned multimodal data to complete the Parkinson's FOG database with expert labels.

Artifacts of each mode were remove separately. EEG data were preprocessed using EEGLab [10] and electrooculogram (EOG) artifacts were removed based on the independent component analysis (ICA) with the average of TP9 and TP10 as the reference. The EEG data was then filtered by a band-pass filter of 0.5-100Hz.

EMG data were filtered by a band-pass filter of 10Hz-500Hz. ACC were filtered by a low-pass filter with a stop frequency of 16Hz. All the data were filtered out the 50 Hz power line interference by a notch filter and normalized.

In the FOG detection, the multimodal FOG data were segmented and assign a common labelled for each segment. The data were segmented using a sliding window method with a window length of 3 seconds and a sliding step size of 0.3 second. Each segment was assigned a common label based on the proportion of FOG time points, that is, the Percentage of FOG (PFG) points defined as (1).

$$PFG = \frac{N_{FOG}}{N_{FOG}+N_N} \times 100\% \qquad (1)$$

where $N_{FOG}$ is the number of FOG data points in the segment, while $N_N$ is the number of FOG-free points. The label of the segment is determined by (2), where $T$ is the appropriate threshold selected by the researcher, which is usually around 0.75-0.85.

$$SegmentLabel = \begin{cases} 1, if\ PFG \geq T \\ -1, if\ PFG < T \end{cases} \qquad (2)$$

The labeling threshold was set to 80% in the following discussion, which means that FOG's appearance in the data segment if over 80% data points were labelled as positive by physicians.

Each segement of the multimodal data and the associated segment-label composed of an effective sample for the following features extraction and classification.

*2.5 Feature extraction*

A total number of sixteen statistical features (as in Table 2) were employed for the classification of FOG according to the results in references. Each unimodal data was used to detect FOG individually and the multimodal data were then used by combining all these features in the detection of FOG. Both intra-subject test error and inter-subject test errors were used to verify the effectiveness of the multimodal data in the detection of FOG.

Table 2 Multimodal features and brief description

| Data | Channel Quantity | Feature | Description |
|---|---|---|---|
| EEG | 25 | $WE_\delta$<br>$WE_\theta$<br>$WE_\alpha$ | Represents changes in energy of FOG and locomotion period of PD patients' EEG signal |
| | | TWE | Represents changes in energy complexity |
| EMG | 3 | MAV | Estimation of the STD of EMG signals |
| | | ZC<br>SSC | Related to the frequency of EMG signals |
| | | WL | Directly related to the EMG signals |
| ACC | 3 | SE | Evaluate the repeatability of the waveform |
| | | STD | Standard Deviation |
| | | TP<br>FI | Detection algorithm of FOG proposed by Moore et al. [13] |

For the EEG mode, 5-scale discrete wavelet transforms (DWT) were applied to each of the 25 channels to obtain five rhythms, that is, the δ wave (0-3.9Hz); θ wave (3.9-7.8Hz); $\alpha$ wave (7.8-15.6Hz); β wave (15.6-31.3Hz); and $\gamma$ wave (31.3-62.5Hz) [14]. The extracted EEG features of each channel were wavelet energy (WE) of a segement of data in δ, θ, $\alpha$-bands and the associated total wavelet entropy (TWE), denoted as $WE_\delta$, $WE_\theta$, $WE_\alpha$, and TWE, respectively. The WE of each component are defined as

$$WE_j = \sum_{k=1}^{N} \left| y_j \right|^2 \quad (3)$$

where $y_j$ denotes the $j^{th}$ components of an EEG channel; $WE_j$ denotes the WE of the $j^{th}$ component of an EEG channel after the DWT; $N$ is the window length of a segment.

The associated TWE is defined as

$$TWE = -\sum_j \frac{WE_j}{\sum_j WE_j} \log \frac{WE_j}{\sum_j WE_j} \quad (4)$$

For the EMG data, four features of single-channel of EMG were extracted including Mean Absolute Value (MAV), Zeros Crossing (ZC), Slope Sign Change (SSC), and Wave Length (WL).

For the ACC data, the accelerations in three directions at lateral tibias of the left or right leg were extracted as the associated Sample Entropy (SE), Standard deviation (STD), Total Power (TP) and Freezing Index (FI) which is defined as the ratio of the power in freezing band (3-8 Hz) and locomotion band (0.5-3 Hz) [13].

*2.6 FOG Detection and Evaluation*

Based on the extracted features, SVM classifier with radial basis function kernel were used to classify FOG from non-FOG locomotion. Four feature combinations were considered, including pure EEG features, pure EMG features, pure ACC features, and multimodal features including all EEG, EMG and ACC features. Simultaneously, divide the dataset into the training set and test set with a quantity ratio of 0.25 randomly, and the method of cross-validation and grid search was used to determine the hyper-parameters of SVM model. Replicating each experiment 20 times, the average value of the criteria, including accuracy, sensitivity, specificity, precision, F1 value, area under curve (AUC) were used to evaluate the classification performance.

The multimodal data FOG detection were evaluated by its discrimination ability of feature combinations. The performance of the different modal data was compared under two different settings, namely, subject-dependent and subject-independent verifications. In the subject-dependent case, the data from each subject were divided into training and test sets. The performance of the multimodal data was evaluated individually for each subject. On the contrary, the subject-independent verification mixed data from all subjects and evaluate the performance of the dataset considering the inter-subject variability.

## 3. Results

### *3.1 Dataset descriptive statistics: Statistical Analysis*

The data have been collected in Beijing Xuanwu Hospital since 2019. Until the paper was written, a total of 18 individuals have been selected based on the inclusion criteria and completed the whole data collection procedures. Among them, data of 12 participants (13 experiments, Patient ID: 08 conduct the experiment twice) are valid and can be used for the investigation of multimodal data FOG detection. (The reasons of the "invalid" data include: 1. Patients' walking posture was affected by wearing the devices due to age reason or leg disease; 2. After analyzing the patient's video, experts concluded that their gait was affected not only by FOG but also other leg diseases; 3. The data were affected by participants' limited stamina, which leads to poor data quality.)

Participants aged between 57 and 81 years (average: 69.1 years), and have disease duration between 1 and 20 years (average: 9.3 years). 10 subjects had conspicuous FOG episodes during the experiments. The total length of data was 222 minutes and 3 seconds. There are 334 FOG events with a total duration of 88 minutes and 19 seconds.

Table 3 Participants' information

| Characteristics | | Average (Total) |
|---|---|---|
| Age | | 69.1±7.9 |
| Disease Duration | | 9.3±6.8 |
| ADL | | 81.3±16.0 |
| FOG-Q | | 16.2±4.2 |
| UPDRS | UPDRS-1 | 10.4±5.5 |
| | UPDRS-2 | 16.3±10.6 |
| | UPDRS-3 | 45.0±16.0 |
| | UPDRS-4 | 2.2±2.9 |
| MMSE | | 28.2±1.5 |
| MOCA | | 23.6±3.6 |
| Length of Data(min:sec) | | 17:05±10:43 (222:03) |
| Total FOG Time(min:sec) | | 06:48±7:36 (88:19) |
| Number of FOG Events | | 25.7±17 (334) |

ADL = Activities of Daily Living Section; UPDRS = Unified Parkinson's Disease Rating Scales; FOG-Q = Freezing of Gait Questionnaire; MMSE = Mini-Mental State Examination, MOCA = Montreal Cognitive Assessment.

### *3.2 Dataset descriptive statistics: Detailed Duration of FOG Events*

In the length of 222 minutes and 3 seconds of valid data, the duration of a single FOG event ranged from 1 to 201s. Over 35% of episodes lasted less than 5s, and over 50% of episodes lasted less than 10s, see Fig.6.

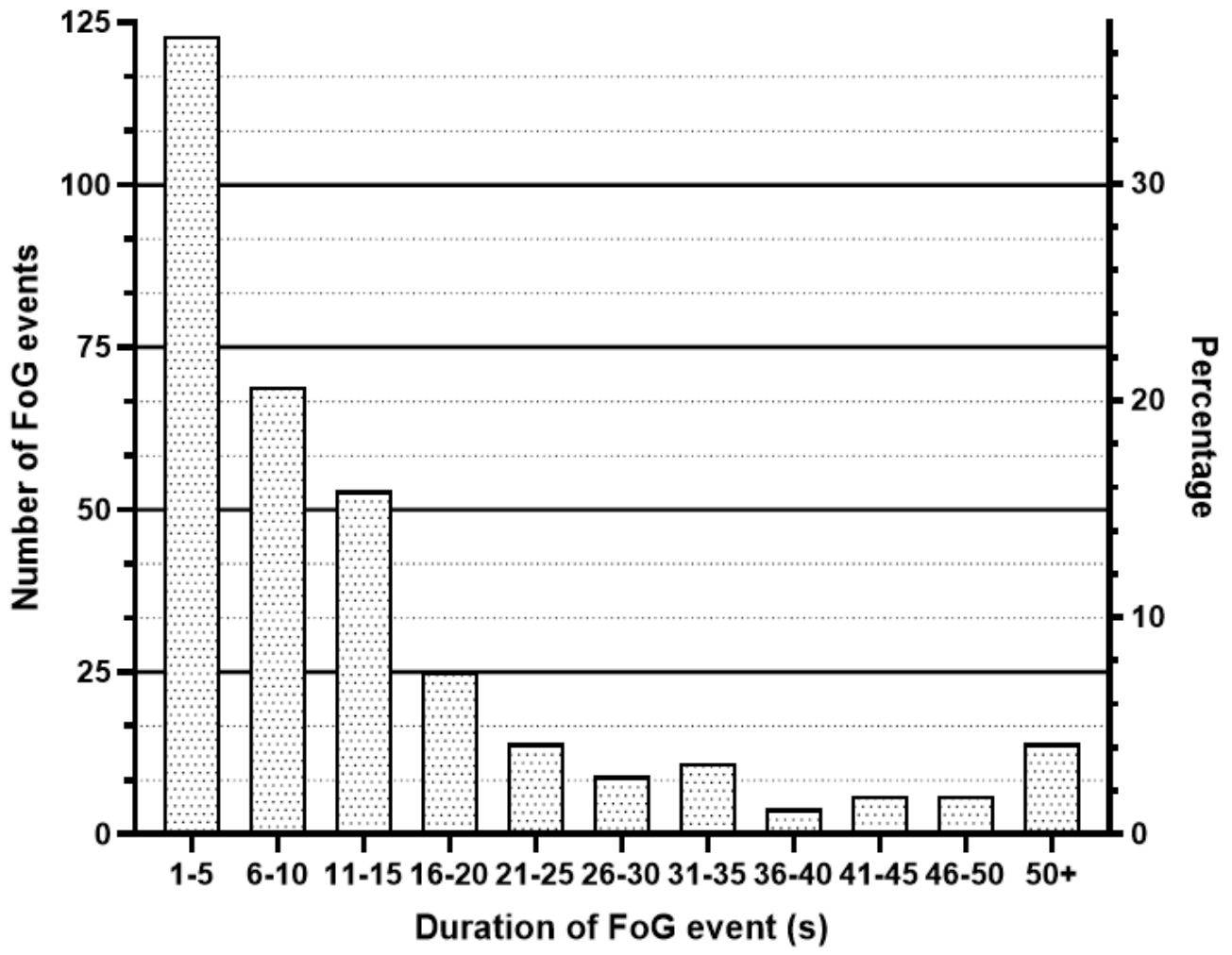

Fig. 6 Distribution of FOG Duration.

Detailed duration of FOG events of each participant is shown in Table 4. It can be observed that the number of FOG events and the duration of each FOG episode vary significantly among 12 subjects. Patient ID:03 suffered a large amount of FOG during data collection while Patient (ID:02) showed a few; most FOG events of Patient ID:09 has a duration less than 5 seconds, while Patient ID:10 have many FOG events that last more than 40 seconds. Such noticeable inter-subject variability in FOG events indicates that results of subject-specific study of FOG detection and prediction can be better that those of subject-independent study.

Table 4 Duration of FOG events

| Patient | Duration (s) | | | | | | | | | | |
|---|---|---|---|---|---|---|---|---|---|---|---|
| ID | 1-5 | 6-10 | 11-15 | 16-20 | 21-25 | 26-30 | 31-35 | 36-40 | 41-45 | 46-50 | 50+ |
| 01 | 6 | 5 | 7 | 1 | 1 | 0 | 2 | 0 | 0 | 0 | 0 |
| 02N | 1 | 0 | 0 | 0 | 0 | 0 | 0 | 0 | 0 | 0 | 0 |
| 03 | 16 | 10 | 6 | 3 | 2 | 0 | 2 | 3 | 0 | 0 | 9 |
| 04 | 5 | 7 | 2 | 0 | 1 | 0 | 0 | 0 | 0 | 0 | 0 |
| 05N | 0 | 0 | 0 | 0 | 0 | 0 | 0 | 0 | 0 | 0 | 0 |
| 06 | 1 | 4 | 5 | 7 | 3 | 1 | 1 | 0 | 0 | 0 | 0 |
| 07 | 10 | 6 | 10 | 5 | 1 | 1 | 0 | 0 | 0 | 0 | 0 |
| 08(1) | 11 | 3 | 2 | 1 | 2 | 2 | 1 | 0 | 0 | 2 | 1 |
| 08(2) | 4 | 6 | 5 | 3 | 0 | 1 | 2 | 1 | 0 | 2 | 0 |
| 09 | 38 | 1 | 0 | 0 | 0 | 0 | 0 | 0 | 0 | 0 | 0 |
| 10 | 4 | 9 | 8 | 1 | 2 | 2 | 0 | 0 | 4 | 1 | 4 |
| 11 | 25 | 16 | 7 | 2 | 1 | 1 | 2 | 0 | 1 | 0 | 0 |
| 12 | 2 | 2 | 1 | 2 | 1 | 1 | 1 | 0 | 1 | 1 | 0 |
| Total | 123 | 69 | 53 | 25 | 14 | 9 | 11 | 4 | 6 | 6 | 14 |

An N next to the patient ID indicates that there almost no FOG appeared during the experiment.
Patient 08 repeated the data collection twice.

### *3.3 Detection of FOG*

Literatures have shown that multimodal information benefits the accurate detection of FOG. In this section, the collected multimodal data were used to compare the performance of each unimodal data in the detection of FOG. Data from difference sources were time-aligned and preprocessed to remove the effect of artifacts. Features were extracted for each single modal data and Support vector machine (SVM) classifiers were trained based on uinmodal sensing data and their combinations.

*3.3.1 Subject-dependent results*

For the subject-specified detection of FOG, the data from each participant were divided into training and test data with a ratio of 3:1. The performance on the test data are reported. The average results over 13 experiments are shown in Fig.7. The results of SVM classifiers shows that the average values of all criteria for all mode combinations exceeded 73%. The average accuracies and AUCs of the four feature combinations exceeded 88% and 90%, respectively. The EEG data performed the best in three unimodal data and EMG performed the worst classification results. The multimodal data which combined all three modes significantly surpassed the performances of single mode features. This indicates that the multimodal data characterised FOG better than single modal data did. However, the EEG produced a comparable results as the mulitmodal data.

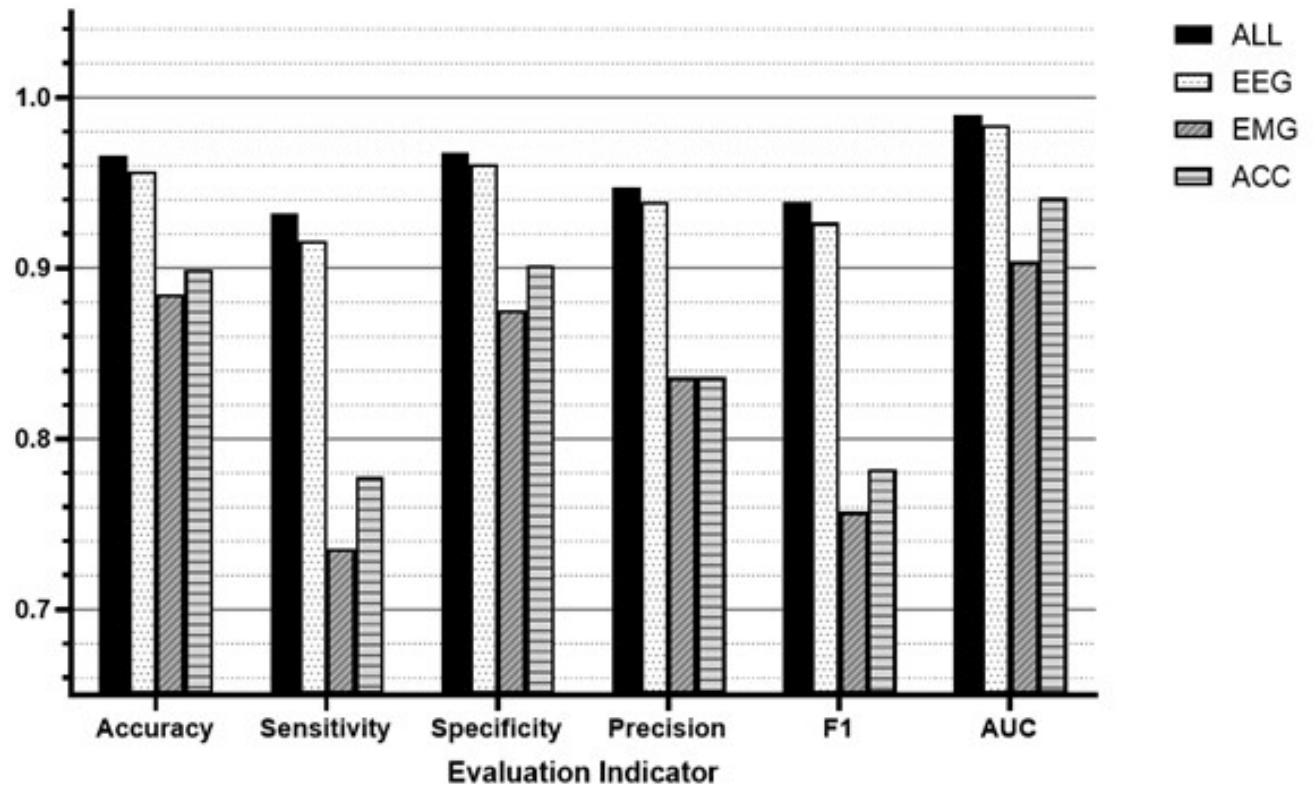

Fig. 7 Average Result of four types of SVM classification in subject-dependent analysis.

Detailed results of the subject-specific analysis were given in Table 6. It can be observed that the duration of FOG events in the data had a significant impact on the classification performance. Either the patient with major FOG ( such as subject ID:03) or patient with minor FOG events (such as subject ID:09) produced relatively poor performance, which may be caused by the unbalanced data in the training of the classifieres.

Table 6 Results of different feature combinations in subject-dependent analysis

| Patient ID | Features | Accuracy | Sensitivity | Specificity | Precision | F1 Value | AUC |
|---|---|---|---|---|---|---|---|
| 01 | EEG | 0.9709 | 0.9420 | 0.9833 | 0.9602 | 0.9510 | 0.9953 |
| | EMG | 0.8866 | 0.8028 | 0.9224 | 0.8162 | 0.8092 | 0.9335 |
| | ACC | 0.8983 | 0.8617 | 0.9136 | 0.8094 | 0.8345 | 0.9613 |
| | ALL | 0.9701 | 0.9380 | 0.9838 | 0.9608 | 0.9492 | 0.9939 |
| 02N | / | / | / | / | / | / | / |
| 03 | EEG | 0.9651 | 0.9855 | 0.8825 | 0.9714 | 0.9784 | 0.9923 |
| | EMG | 0.8303 | 0.9589 | 0.3093 | 0.8492 | 0.9007 | 0.7828 |
| | ACC | 0.9374 | 0.1154 | 0.9953 | 0.7438 | 0.1743 | 0.9007 |
| | ALL | 0.9781 | 0.7570 | 0.9938 | 0.8989 | 0.8204 | 0.9817 |
| 04 | EEG | 0.9784 | 0.9462 | 0.9850 | 0.9292 | 0.9375 | 0.9941 |
| | EMG | 0.9448 | 0.8057 | 0.9729 | 0.8570 | 0.8301 | 0.9616 |
| | ACC | 0.9400 | 0.7500 | 0.9799 | 0.8867 | 0.8117 | 0.9622 |
| | ALL | 0.9770 | 0.9304 | 0.9864 | 0.9326 | 0.9314 | 0.9958 |
| 05N | / | / | / | / | / | / | / |
| 06 | EEG | 0.9564 | 0.9136 | 0.9736 | 0.9331 | 0.9232 | 0.9847 |
| | EMG | 0.8816 | 0.7853 | 0.9196 | 0.7938 | 0.7894 | 0.9269 |
| | ACC | 0.8575 | 0.7846 | 0.8863 | 0.7313 | 0.7566 | 0.9164 |
| | ALL | 0.9653 | 0.9497 | 0.9714 | 0.9293 | 0.9393 | 0.9907 |
| 07 | EEG | 0.8789 | 0.8646 | 0.8895 | 0.8525 | 0.8585 | 0.9423 |
| | EMG | 0.8555 | 0.7897 | 0.9042 | 0.8591 | 0.8229 | 0.9220 |
| | ACC | 0.8709 | 0.8283 | 0.9024 | 0.8624 | 0.8450 | 0.9349 |
| | ALL | 0.9252 | 0.9484 | 0.9080 | 0.8845 | 0.9151 | 0.9693 |
| 08(1) | EEG | 0.9600 | 0.9439 | 0.9719 | 0.9613 | 0.9525 | 0.9878 |
| | EMG | 0.8872 | 0.8694 | 0.9005 | 0.8661 | 0.8677 | 0.9501 |
| | ACC | 0.9353 | 0.9289 | 0.9400 | 0.9198 | 0.9242 | 0.9809 |
| | ALL | 0.9715 | 0.9608 | 0.9794 | 0.9718 | 0.9663 | 0.9895 |
| 08(2) | EEG | 0.9459 | 0.9487 | 0.9432 | 0.9404 | 0.9445 | 0.9766 |
| | EMG | 0.8874 | 0.8743 | 0.8999 | 0.8932 | 0.8836 | 0.9368 |
| | ACC | 0.9318 | 0.9533 | 0.9111 | 0.9114 | 0.9318 | 0.9799 |
| | ALL | 0.9560 | 0.9587 | 0.9534 | 0.9516 | 0.9551 | 0.9857 |
| 09 | EEG | 0.9762 | 0.7429 | 0.9928 | 0.8814 | 0.8047 | 0.9801 |
| | EMG | 0.9357 | 0.0545 | 0.9980 | 0.7123 | 0.1013 | 0.7856 |
| | ACC | 0.9238 | 0.9733 | 0.7198 | 0.9348 | 0.9536 | 0.9519 |
| | ALL | 0.9718 | 0.9863 | 0.9130 | 0.9787 | 0.9825 | 0.9945 |
| 10 | EEG | 0.9422 | 0.9042 | 0.9720 | 0.9620 | 0.9322 | 0.9827 |
| | EMG | 0.8772 | 0.8414 | 0.9054 | 0.8744 | 0.8575 | 0.9372 |
| | ACC | 0.8327 | 0.8507 | 0.8186 | 0.7860 | 0.8170 | 0.9033 |
| | ALL | 0.9585 | 0.9301 | 0.9806 | 0.9741 | 0.9516 | 0.9889 |
| 11 | EEG | 0.9639 | 0.9284 | 0.9826 | 0.9657 | 0.9466 | 0.9940 |
| | EMG | 0.8600 | 0.7292 | 0.9288 | 0.8444 | 0.7821 | 0.9146 |
| | ACC | 0.8490 | 0.7699 | 0.8907 | 0.7879 | 0.7787 | 0.9173 |
| | ALL | 0.9666 | 0.9337 | 0.9839 | 0.9684 | 0.9507 | 0.9944 |
| 12 | EEG | 0.9852 | 0.9566 | 0.9927 | 0.9718 | 0.9640 | 0.9947 |
| | EMG | 0.8888 | 0.5809 | 0.9693 | 0.8333 | 0.6836 | 0.8905 |
| | ACC | 0.9125 | 0.7349 | 0.9589 | 0.8240 | 0.7767 | 0.9427 |
| | ALL | 0.9857 | 0.9637 | 0.9915 | 0.9674 | 0.9655 | 0.9982 |
| Average | EEG | 0.95664 | 0.91606 | 0.96081 | 0.93900 | 0.92666 | 0.98404 |
| | EMG | 0.88501 | 0.73564 | 0.87547 | 0.83626 | 0.75709 | 0.90378 |
| | ACC | 0.89901 | 0.77738 | 0.90150 | 0.83613 | 0.78219 | 0.94106 |
| | ALL | 0.96597 | 0.93244 | 0.96774 | 0.94709 | 0.93882 | 0.98932 |

An N next to the patient ID indicates that there almost no FOG appeared during the experiment.
ALL = All features together.

### *3.3.2 Subject-independent results*

Subject-independent study has also been conduced to verify the effects of inter-subject variability, where the data from all subjects are mixed and divided into training and test data with the same ratio as in subject-specific study. The same set of features and SVM classifier with same settings were used to discriminate the FOG from normal locomotion. This time only the multimodal data were studied. Repeat the classification 20 times and the average value of the evaluation indicator on the test set is shown in Table 5.

Table 5 Average result of subject-independent verification based on multimodal data

| Evaluation Indicator | Average Result |
|---|---|
| Accuracy | 0.9625±0.0018 |
| Sensitivity | 0.9487±0.0025 |
| Specificity | 0.9722±0.0020 |
| Precision | 0.9605±0.0026 |
| F1 Value | 0.9604±0.0022 |
| AUC | 0.9546±0.0008 |

The box plot of each criterion is shown in Fig 8.

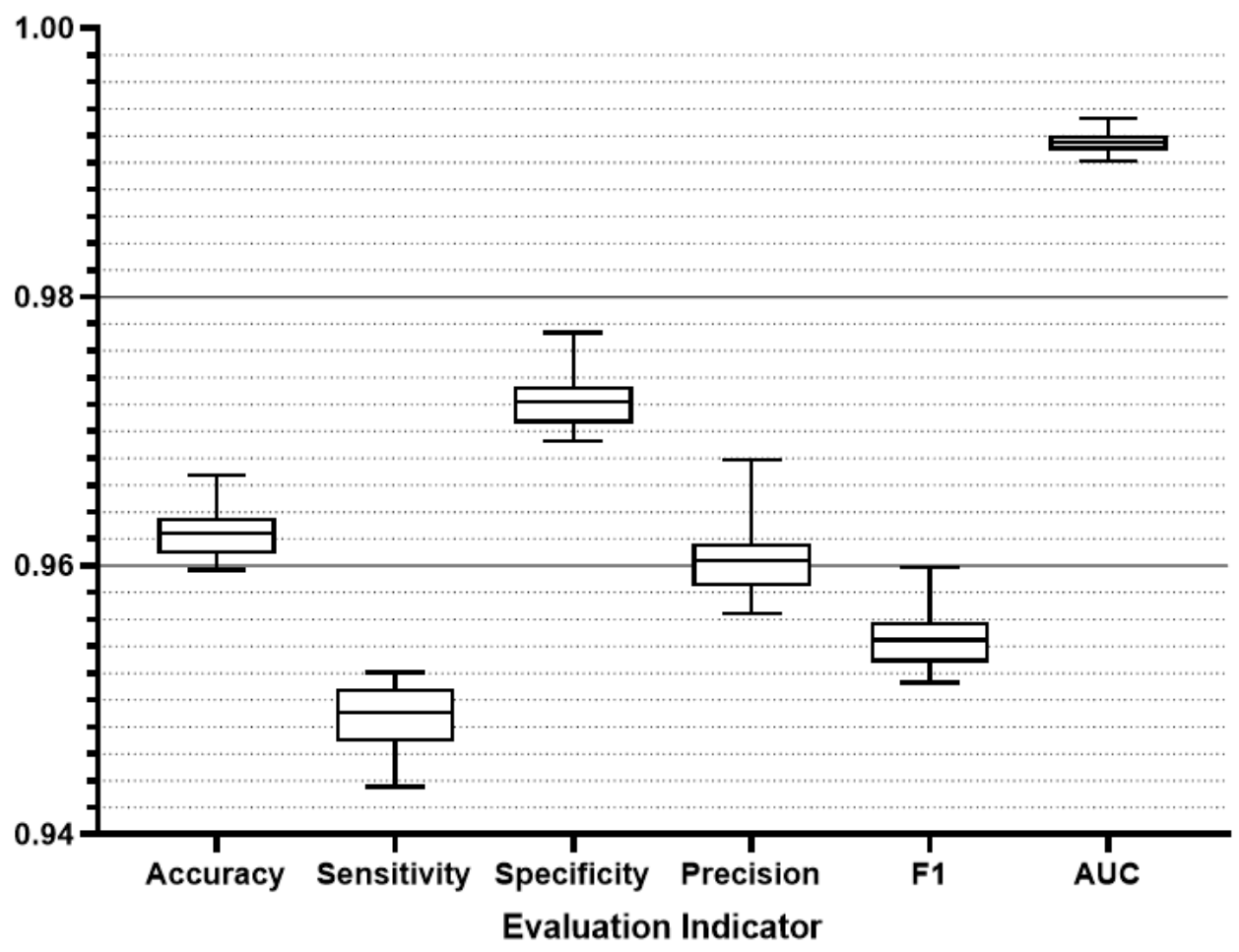

Fig 8 Result of SVM classification in subject-independent method.

The results showed that all five metrics were averagely greater than 0.94 with small variances. This indicates that the multimodal FOG dataset are of a high quality and have good generalization ability based on the extracted multimodal features.

**4 Conclusions**

FOG is a very disabiling symptom of PD, affecting about 50% of PD patients and 80% advanced PD patients [15]. Studies have showed that FOG is not only a pure motor phenomenon but related to a complex interplay between motor, cognitive and affective facgors. A full characterisation of FOG is crucial for FOG detection/prediction and prompt intervention.

A multimodal data sensoring system, including EEG, ACC, EMG and SC have been designed for the acquisition of the multimodal FOG data. A pipeline for the detection of FOG based on multimodal data has been proposed, including the experiment design, data acquisition, data pre-processing, feature extraction, classification and performance evaluation. Combinging all multimodal data benefits the detection of FOG and produced a consistantly better results than single modal data.

The discrimination abilities of each mode have also been compared and results showed that the EEG signals has the best performance in the detection of FOG than ACC and EMG did. This can be because the EEG signals used more features than other modes. However, the preparation, acquisition and preprocessing of EEG data can be costly and time-consuming. Therefore, EEG is not suitful for long-term monitoring of FOG in a living condition even though it has the best performance. It is worthy to explpore the dynamic depedence among the multimodal data and develop an easy-to-implement long-term FOG monitoring method. This will be one of our future works.

**Appendix A: Rule of data alignment**

The experiments complied with the following rules in order to simplify the data alignment and annotation:

1) Each acquisition subsystem has its own millisecond timer, and the data are parallelly stored using their own timestamps. The start and end time of each task were recorded by a separate stopwatch, which was calibrated with the world time, as the world time of each task;
2) All sensors simultaneously sampled data during the whole experiment;
3) The entire process of the experiment was recorded as a video for physicians to label the FOG episodes;
4) All sampling starts at least 30 seconds earlier than each task's kickoff (the patient was asked to quickly stand up and sit down three times of which the sharp changing ACC was used as the task start instant in the ACC data);
5) Assign the start and end time which recorded by the stopwatch to each of the multimodal data and also the video to callibrate the timestamps.

## Appendix B: Data and code availability

Both the raw data which directly obtained from the hardware system, and the filtered data which have be preprocessed and labeled, are published on *Mendeley Data*. The specific description of dataset can be referred to supplementary materials or online dataset's files. Raw data available at [12], filtered data available at [11].

The preprocessing of EEG data were conducted with EEGLab. The codes of the preprocessing of EMG and ACC, labeling of raw data, and of feature extraction, are available with the data file [11].

## Author contributions

YG, WZ, LW, HL, and YW contributed to the design of the study. WZ, LW, HL, KP, LM, HF, and JP helped with different aspects related to the implementation of the research protocol including data collection. HL and DH helped with data analysis and processing. HL, YG and DH wrote the manuscript. All authors provided editorial feedback and contributed to the final approval of the manuscript.

## Acknowledgements

Authors YG, HL, DH, LW, YW, and KP gratefully acknowledge the support from the National Natural Science Foundation of China (Grant No. 61876015), the Beijing Natural Science Foundation, China (Grant No. 4202040), and Science and Technology Innovation 2030 Major Program of China (Grant No. 2018AAA001400). Authors WZ, LM, HF, JP and PC acknowledge the partial support from Beijing Municipal Administration of Hospitals' Mission Plan (No. SML20150803), The National Key R&D Program of China (No. 2018YFC1312001, 2017YFC0840105), Beijing Municipal Science & Technology Commission (No. Z171100000117013).